\documentclass[aps,pra,reprint,amsfonts,amssymb,amsmath]{revtex4-1}
\usepackage{graphicx}
\usepackage{epstopdf}

\usepackage{dcolumn}
\usepackage{bm}
\usepackage{bigints}
\usepackage{xcolor}  
\usepackage{subdepth} 
\usepackage{hyperref} 
\usepackage{braket}
\hypersetup{colorlinks,
    linkcolor={blue},
    citecolor={blue},
    urlcolor=blue}

\begin{document}

\title{Critical phenomena and nonlinear dynamics in a spin ensemble strongly \\coupled to a cavity. II. Semiclassical-to-quantum boundary}
\author{Matthias Zens}
\email[matthias.zens@tuwien.ac.at]{}
\author{Dmitry O. Krimer}
\author{Stefan Rotter}
\affiliation{Institute for Theoretical Physics, Vienna University of Technology (TU Wien), Wiedner Hauptstra\ss e 8-10/136, A--1040 Vienna, Austria, EU}

\begin{abstract}
We numerically study the dynamics and stationary states of a spin ensemble strongly coupled to a single-mode resonator subjected to loss and external driving. Employing a generalized cumulant expansion approach we analyze finite-size corrections to a semiclassical description of amplitude bistability, which is a paradigm example of a driven-dissipative phase transition. Our theoretical model allows us to include inhomogeneous broadening of the spin ensemble and to capture in which way the quantum corrections approach the semiclassical limit for increasing ensemble size $N$. We set up a criterion for the validity of the Maxwell-Bloch equations and show that close to the critical point of amplitude bistability even very large spin ensembles consisting of up to $10^4$ spins feature significant deviations from the semiclassical theory.   
\end{abstract}

\maketitle

\section{Introduction}
The dynamics of open many-body quantum systems is of fundamental importance for various branches of physics \cite{Daley2014}. In particular, so-called hybrid quantum systems, whose technological relevance requires them to be open, have shifted to the center of attention over the last decade \cite{Xiang2012a,Kurizki2015}. As a prominent example, spin ensembles coupled to a cavity mode emerged as a powerful platform for quantum computation \cite{Tordrup2008, Wesenberg2009,  Ping2012}, quantum memories \cite{Simon2010a, Kubo2012b, Grezes2016}, and quantum communication \cite{Duan2001, Kimble2008}. Beside their technological significance, the driven and dissipative character of spin-cavity systems makes them also well-suited to study fundamental aspects of non-equilibrium many-body physics \cite{Houck2012, Tomadin2010, Buchhold2013}. Corresponding theoretical descriptions clearly profit from the fact that the interactions among the individual spins or atoms are mediated via few common cavity modes only. This leads to extremely long-ranged interactions and suppressed fluctuations \cite{Buchhold2013, Foss-Feig2017}, often enabling accurate semiclassical descriptions of these systems.

Among semiclassical approaches the seminal Maxwell-Bloch equations play a distinguished role in quantum optics. Being based on neglecting the correlations between spins and the electromagnetic field they successfully describe many effects of lasers \cite{Haken1985}, superradiance \cite{Dicke1954a,Svidzinsky2015,Rose2017a}, critical slowing down \cite{Bonifacio1979,Angerer2017,Krimer2019}, and amplitude bistability \cite{Gibbs1976,Bonifacio1976,Bonifacio1978a,Lugiato1984,Brennecke2008,Angerer2017,Krimer2019}. In a separate companion paper we use the Maxwell-Bloch equations to study the dynamics of macroscopic spin ensembles near the critical point of amplitude bistability and analyze the effect of critical slowing down in the presence of inhomogeneous broadening \cite{Krimer2019}. Here, we will investigate to what extent the Maxwell-Bloch equations themselves are well-justified to describe the actual quantum dynamics of such systems on experimentally relevant time scales. This question is of particular interest near the critical point of a driven-dissipative phase transition as in the case of optical bistability. Here the semiclassical Maxwell-Bloch equations give rise to two stable steady state solutions combined with a hysteresis effect, while a full quantum treatment predicts a unique steady state that can deviate dramatically from semiclassical results \cite{Lugiato1984}. One major source of this deviation is the bimodal character of the quantum-mechanical probability distribution, which leads to a switching of the system between the two semiclassically stable branches \cite{Dombi2013, Fink2017,Casteels2017}. Since, however, the switching time between the stable solutions diverges with the system size \cite{Fink2017,Casteels2017}, the Maxwell-Bloch equations describe well how large systems evolve dynamically for experimentally observable time scales. The natural question to ask in this specific context is thus how many spins are required to constitute an ensemble that is sufficiently ''large'' to enter the thermodynamic limit where the semiclassical solutions are sufficiently accurate to describe the dynamics of this quantum system close to the bistability region \cite{Rempe1991a,Dombi2013}.

To carry out such a comparison between the semi-classical and the quantum dynamics, the time evolution of a finite number of spins inside a cavity needs to be evaluated \cite{Sarkar1987,Rempe1991a,Dombi2013,Fink2017,Shirai2018,Carmichael_QO2008,Dhar2018,Kirton2017,Mori2013}. Due to computational constraints, a full quantum mechanical treatment of cavity-spin systems is, however, limited to small spin ensembles or to few excitations in the system \cite{Carmichael_QO2008}. 
Results for spin ensembles of up to $8$ spins \cite{Dombi2013} showed a rapid convergence of the quantum case to the semiclassical limit, but failed to provide a quantitative analysis of the corresponding boundary. Also further extensive theoretical efforts to describe spin-cavity systems of increasing ensemble size on a full quantum level \cite{Kirton2017,Shirai2018,Dhar2018} left the transition between the semiclassical and the full quantum case mostly uncharted. Since these two realms occupy opposing limits with respect to the system size, special techniques are required to bridge the gap between the microscopic and the macroscopic domain.

For a closed system of completely symmetric spin ensembles, e.g., the scaling of quantum corrections as a function of the number of two-level systems was studied using a WKB approach analyzing the eigenvalue spectrum \cite{Keeling2009a}. In the present paper we study the onset of quantum corrections in the complementary setting of an open system with external driving and dissipation. Using a generalized cumulant expansion approach \cite{Kubo1962a,Leymann2013,Henschel2010,Leymann2014,Kreamer2015,Vardi2001b,Vardi2001c,Casteels2016}, we establish a criterion for the validity of the semiclassical Maxwell-Bloch equations for a wide range of parameters taking the effect of inhomogeneous broadening explicitly into account. 

Our paper is organized as follows: In Sec.~\ref{sec:model} we introduce the model and derive the hierarchic set of equations of motion for expectation values (Sec.~\ref{sec:EoM}). We review the semiclassical approximation (Sec.~\ref{sec:SC}) and present a generalized truncation scheme based on higher orders of cumulants (Sec.~\ref{sec:CE}). The effect of amplitude bistability is investigated in Sec.~\ref{sec:results} starting with an homogeneous spin ensemble (Sec.~\ref{sec:Hom}) and examining the semiclassical-to-quantum boundary (Sec.~\ref{sec:SCQboundary}). The presented analysis is then extended to the case of an inhomogeneously broadened spin ensemble (Sec.~\ref{sec:inHom}). In Sec.~\ref{sec:conclusions} we draw our conclusions.
\section{Model}
\label{sec:model}
The system we consider consists of $N$ two-level atoms or spins with transition frequencies $\omega_{j}$ coupled  to a single-mode cavity with coupling strength $g_j$. The two-level emitters may or may not exhibit inhomogeneous broadening or coupling, both of which can be treated with the model at hand. The cavity is coherently driven by an external field of strength $\eta$ and frequency $\omega_p$.
The starting point for the theoretical model is the Tavis-Cummings Hamiltonian \citep{TavisCummings}, which, in a rotating frame with driving frequency $\omega_p$, reads ($\hbar=1$)
\begin{multline}
\mathcal{H}=\Delta_c\,a^\dag a +\frac{1}{2}\sum_{j=1}^N\Delta_j \sigma_j^z
+\sum_{j=1}^N[g_j\sigma_j^-a^\dag+g_j^*\sigma_j^+a]\\
+i[\eta(t) a^\dag-\eta^*(t) a],
\label{eq:Hamiltonian}
\end{multline}
where $\Delta_c\equiv \omega_c-\omega_p$ and $\Delta_j\equiv\omega_j-\omega_p$ are the detunings of the cavity frequency $\omega_c$ and of the individual spin frequencies $\omega_j$ with respect to the external driving field of frequency $\omega_p$.
Here $a^\dag$ and $a$ are the creation and annihilation operators of the single cavity mode and $\sigma_j^z$, $\sigma_j^+$,  and $\sigma_j^-$ are the Pauli operators corresponding to the individual spins. In the following we consider the driving amplitude to be constant ($\eta(t)=\eta$) and without loss of generality assume $\eta^*=\eta$ as well as $g^*_j=g_j$.

The dissipation of the system is described by the Lindblad superoperator
\begin{align}
\mathcal{ L}_D(\rho)=& \,\kappa\,(2a\rho a^\dag-a^\dag a\,\rho-\rho\, a^\dag a)+\gamma_p\sum\limits_{j=1}^N(\sigma_j^z\rho\,\sigma_j^z-\rho\,)\notag\\
&+\gamma_h\sum\limits_{j=1}^N(2\sigma_j^-\rho\,\sigma_j^+-\sigma_j^+\sigma_j^-\rho-\rho\,\sigma_j^+\sigma_j^-),
\label{eq:Lindbladian}
\end{align}
where the first term corresponds to the cavity loss at rate $\kappa$; the second term gives non-radiative dephasing of the individual spins with rate $\gamma_p/2$, and the last term describes their radiative decay with rate $\gamma_h$.

The total dynamics of the driven dissipative system is then given by the master equation 
\begin{equation}
\frac{d}{dt}\rho=\frac{1}{i}[\mathcal{H},\rho]+\mathcal{ L}_D(\rho),
\label{eq:masterequation}
\end{equation} 
with $\mathcal{H}$ and $\mathcal{ L}_D  $ given by Eq.~\eqref{eq:Hamiltonian} and \eqref{eq:Lindbladian}, respectively, and $\rho$ being the density operator of the total cavity-spin system. 

\subsection{Equations of motion for expectation values}
\label{sec:EoM}
Full quantum solutions for the density operator of the system or of chosen subsystems can be obtained via direct integration of the master equation \cite{Carmichael1991, Auffeves2011, Saez-Blazquez2017}, quantum trajectory methods \cite{Daley2014, Gardiner1992, Vukics2007, Dombi2013,Fink2017} or, as recently shown in \citep{Dhar2018} also by variational renormalization group methods. All of these approaches are limited, however, in the number of spins or excitations in the system. Since we are interested in the relation between quantum mechanical and semiclassical solutions over a wide range of parameters, we take a different approach here and directly solve for the expectation values of the operators of interest \cite{Leymann2014}. Multiplying the master equation \eqref{eq:masterequation} with the given operator, taking the trace operation and using the cyclic permutation of the trace, it is straight-forward to obtain the following equations of motion (EoM) for the expectation values $\langle a\rangle$, $\langle\sigma_j^-\rangle$, and $\langle\sigma_j^z\rangle$:
\begin{align} 
\label{eq:hierarchya}
&\frac{d}{dt}\langle a\rangle =-(\kappa+i\,\Delta_c)\langle a\rangle-i\,\sum_{j=1}^Ng_j\langle\sigma_j^-\rangle+\eta\,, \\ 
&\frac{d}{dt}\langle \sigma_j^-\rangle =-(\gamma_h+2\gamma_p+i\,\Delta_j)\langle\sigma_j^-\rangle+i\,g_j\langle\sigma_j^za\rangle\,,\\[2mm]
&\frac{d}{dt}\langle\sigma_j^z\rangle =-2\gamma_h(\langle\sigma_j^z\rangle+1)
+2i\,g_j(\langle\sigma_j^-a^\dag\rangle-\langle\sigma_j^+a\rangle). 
\label{eq:hierarchyc}
\end{align}  

In the following we denote expectation values involving a product of \textit{n} operators as nth order expectation values. The interaction part of the Hamiltonian couples EoM for \textit{n}th-order expectation values to EoM for (\textit{n}$+1$)th order expectation values and thereby creates an infinite hierarchy of coupled equations. (The EoM up to third-order expectation values are shown in the Appendix.) To solve the dynamics of the system, the hierarchy of equations thus needs to be truncated at some level. In the following we use a truncation procedure based on a cumulant expansion \cite{Kubo1962a,Henschel2010,Leymann2013,Leymann2014,Kreamer2015} to obtain a closed set of equations that can be solved numerically.

\subsection{Semiclassical approximation}
\label{sec:SC}
The most prominent approach for driven dissipative spin-cavity systems of the type described in the previous section is to solve Eqs.~\eqref{eq:hierarchya}-\eqref{eq:hierarchyc} in the semiclassical limit by applying a self-consistent field approximation \citep{Bonifacio1978a},
\begin{align}
\label{eq:fullfacta}
\langle\sigma_j^za\rangle&\approx\langle\sigma_j^z\rangle\langle a\rangle,\\
\label{eq:fullfactb}
\langle\sigma_j^-a^\dag\rangle&\approx\langle\sigma_j^-\rangle\langle a^\dag\rangle. 
\end{align}
Using this full factorization, the hierarchy of equations \eqref{eq:hierarchya}-\eqref{eq:hierarchyc} truncates at the 1st order according to our previous notation and one derives the seminal Maxwell-Bloch equations:
\begin{align} 
\label{eq:MaxwellBlocha}
&\frac{d}{dt}\langle a\rangle =-(\kappa+i\,\Delta_c)\langle a\rangle-i\,\sum_{k=1}^Ng_j\langle\sigma_j^-\rangle+\eta\,, \\ 
&\frac{d}{dt}\langle \sigma_j^-\rangle =-(\gamma_h+2\gamma_p+i\,\Delta_j)\langle\sigma_j^-\rangle+i\,g_j\langle\sigma_j^z\rangle\langle a\rangle\,,\\[2mm]
&\frac{d}{dt}\langle\sigma_j^z\rangle =-2\gamma_h(\langle\sigma_j^z\rangle+1)
+2i\,g_j(\langle\sigma_j^-\rangle\langle a^\dag\rangle-\textit{c.c.})\,, 
\label{eq:MaxwellBlochb}
\end{align}  
where \textit{c.c.} stands for the complex conjugate of the previous term.
The full factorization of expectation values employed above is well justified for an infinite number of spins, since fluctuations in the large $N$ limit decrease with $1/N$ \cite{Mori2013,Carmichael_QO2008}. An important property of the semiclassical limit is that with the transformations
\begin{equation}
\label{eq:scaling}
\braket{a}\to\braket{a}/\sqrt{N},\quad g_j \to g_j/\sqrt{N},\quad\eta\to\eta\sqrt{N},
\end{equation} and fixed spin distribution $\rho(\omega_j)$, the governing equations are invariant under variation of the number of spins $N$ inside the ensemble. Any deviations from the scaling given by Eq.~\eqref{eq:scaling} can therefore be attributed to quantum corrections to the semiclassical equations \citep{Dombi2013}. 

\subsection{Cumulant expansion approach}
\label{sec:CE}
The basis of the semiclassical limit is the full factorization given by Eqs.~\eqref{eq:fullfacta}-\eqref{eq:fullfactb}, neglecting the second order cumulant, which for two arbitrary operators $\hat{A}$ and $\hat{B}$ is defined as
\begin{equation}
\braket{\hat{A}\,\hat{B}}_c=\braket{\hat{A}\,\hat{B}}-\braket{\hat{A}}\braket{\hat{B}}.
\end{equation} 
To include correlations between operators in the model it is necessary to keep the EoM for expectation values of higher order explicitly and truncate the hierarchy of equations at some higher level. The explicit formulas for third- and fourth-order cumulants read \cite{Kubo1962a}
\begin{flalign} 
\braket{\hat{A}\hat{B}\hat{C}}_c=&\braket{\hat{A}\hat{B}\hat{C}}-\braket{\hat{A}\hat{B}}\braket{\hat{C}}-\braket{\hat{A}\hat{C}}\braket{\hat{B}}
-\braket{\hat{B}\hat{C}}\braket{\hat{A}}\notag\\&+2\braket{\hat{A}}\braket{\hat{B}}\braket{\hat{C}}\,,&
\label{eq:cumulant 3rd order}
\end{flalign} 
\begin{flalign} 
\braket{\hat{A}\hat{B}\hat{C}\hat{D}}_c&=\braket{\hat{A}\hat{B}\hat{C}\hat{D}}-\Bigl(\braket{\hat{A}}\braket{\hat{B}\hat{C}\hat{D}}+\braket{\hat{B}}\braket{\hat{A}\hat{C}\hat{D}}
\notag\\
&+\braket{\hat{C}}\braket{\hat{A}\hat{B}\hat{D}}+\braket{\hat{D}}\braket{\hat{A}\hat{B}\hat{C}}+\braket{\hat{A}\hat{B}}\braket{\hat{C}\hat{D}}
\notag\\
&+\braket{\hat{A}\hat{C}}\braket{\hat{B}\hat{D}}+\braket{\hat{A}\hat{D}}\braket{\hat{B}\hat{C}}\Bigr)&\notag\\
&+2\Bigl(\braket{\hat{A}\hat{B}}\braket{\hat{C}}\braket{\hat{D}}+\braket{\hat{A}\hat{C}}\braket{\hat{B}}\braket{\hat{D}}
\notag\\
&+\braket{\hat{A}\hat{D}}\braket{\hat{B}}\braket{\hat{C}}+\braket{\hat{B}\hat{C}}\braket{\hat{A}}\braket{\hat{D}}\notag\\
&+\braket{\hat{B}\hat{D}}\braket{\hat{A}}\braket{\hat{C}}
+\braket{\hat{C}\hat{D}}\braket{\hat{A}}\braket{\hat{B}}\Bigr)&\notag\\
&-6\braket{\hat{A}}\braket{\hat{B}}\braket{\hat{C}}\braket{\hat{D}}\,.&\label{eq:cumulant 4th order}
\end{flalign} 
The expressions for cumulants of high order are quite cumbersome but they allow us to represent an expectation value of a given order by a cumulant of the same order and by expectation values of lower order only. Therefore a closed system of equations can be obtained by neglecting higher orders of cumulants. In the following we call a truncation scheme that keeps EoM of expectation values up to \textit{n}th-order but neglects cumulants of the (\textit{n}$+1$)th-order as the \textit{n}th-order cumulant expansion (CE\textit{n}). The semiclassical Maxwell-Bloch equations, hence, can be considered as a first-order cumulant expansion (CE1).

For the present paper we apply a cumulant expansion of second (CE2) and third order (CE3) to the coupled EoM of the driven dissipative spin system (Appendix). The CE2 then consists of 12 coupled EoM for the expectation values: $
\braket{a}$, $\braket{\sigma_k^-}$, $\braket{\sigma_k^z}$, $\braket{\sigma_k^za}$, $\braket{\sigma_k^z\sigma_j^-}$, $\braket{\sigma_k^-a^\dag}$, $\braket{\sigma_k^+\sigma_j^-}$, $\braket{\sigma_k^-a}$, $\braket{a^\dag a^\dag}$, $\braket{a^\dag a}$, $\braket{\sigma_k^z\sigma_j^z}$, and $\braket{\sigma_k^-\sigma_j^-}$. All third-order expectation values are expanded according to
 \begin{flalign} 
\braket{\hat{A}\hat{B}\hat{C}}\approx &\;\braket{\hat{A}\hat{B}}\braket{\hat{C}}+\braket{\hat{A}\hat{C}}\braket{\hat{B}}
+\braket{\hat{B}\hat{C}}\braket{\hat{A}}\notag\\&-2\braket{\hat{A}}\braket{\hat{B}}\braket{\hat{C}}\,,&
\label{eq:truncation 3rd order}
\end{flalign} 
where third-order cumulants $\braket{\hat{A}\hat{B}\hat{C}}_c$ are neglected.
For the CE3 we eliminate this approximation and extend the EoM by $13$ additional equations for the expectation values $\braket{\sigma_k^za^\dag a}$, $\braket{\sigma_k^-a^\dag a}$, $\braket{\sigma_k^-a^\dag a^\dag}$, $\braket{\sigma_k^za a}$, $\braket{\sigma_k^-a a}$, $\braket{a^\dag a a}$, $\braket{a a a}$, $\braket{\sigma_k^z\sigma_j^z a}$, $\braket{\sigma_k^-\sigma_j^- a^\dag}$, $\braket{\sigma_k^+\sigma_j^- a}$, $\braket{\sigma_k^z\sigma_j^- a^\dag}$, $\braket{\sigma_k^z\sigma_j^- a}$, and $\braket{\sigma_k^-\sigma_j^- a}$. Note that third-order expectation values containing only spin operators are not included and are truncated on the level of Eq.~\eqref{eq:truncation 3rd order}, which is justified since correlations among three spins play only a minor role as compared to correlations between spins and the collective cavity mode. All fourth-order expectation values that show up in the EoM of the CE3 are expanded as   
\begin{flalign} 
\braket{\hat{A}\hat{B}\hat{C}\hat{D}}&\approx\braket{\hat{A}}\braket{\hat{B}\hat{C}\hat{D}}+\braket{\hat{B}}\braket{\hat{A}\hat{C}\hat{D}}
+\braket{\hat{C}}\braket{\hat{A}\hat{B}\hat{D}}\notag\\
&+\braket{\hat{D}}\braket{\hat{A}\hat{B}\hat{C}}+\braket{\hat{A}\hat{B}}\braket{\hat{C}\hat{D}}
+\braket{\hat{A}\hat{C}}\braket{\hat{B}\hat{D}}\notag\\
&+\braket{\hat{A}\hat{D}}\braket{\hat{B}\hat{C}}-2\Bigl(\braket{\hat{A}\hat{B}}\braket{\hat{C}}\braket{\hat{D}}\notag\\
&+\braket{\hat{A}\hat{C}}\braket{\hat{B}}\braket{\hat{D}}
+\braket{\hat{A}\hat{D}}\braket{\hat{B}}\braket{\hat{C}}\notag\\
&+\braket{\hat{B}\hat{C}}\braket{\hat{A}}\braket{\hat{D}}+\braket{\hat{B}\hat{D}}\braket{\hat{A}}\braket{\hat{C}}
\notag\\
&+\braket{\hat{C}\hat{D}}\braket{\hat{A}}\braket{\hat{B}}\Bigr)
+6\braket{\hat{A}}\braket{\hat{B}}\braket{\hat{C}}\braket{\hat{D}}\,,&
\label{eq:truncation 4th order}
\end{flalign} 
with fourth-order cumulants $\braket{\hat{A}\hat{B}\hat{C}\hat{D}}_c$ being neglected. The closed set of equations resulting from the CE2 and CE3 is solved numerically to obtain the dynamics and stationary states of the driven-dissipative spin system.

\section{amplitude bistability}
\label{sec:results}
%
%
\begin{figure}
\includegraphics[angle=0,width=1.\columnwidth]{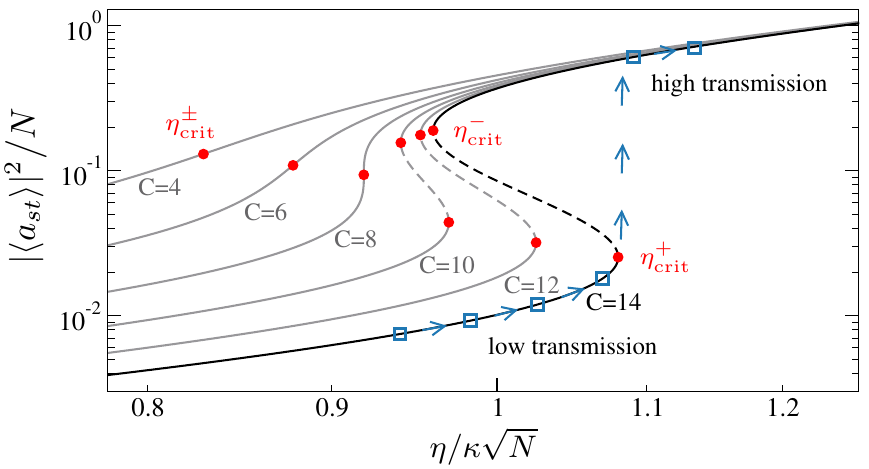}
\vspace*{-0.6cm}
\caption{Stationary solutions for the cavity probability amplitude $|\langle a_{st}\rangle|^2$ as a function of the driving amplitude $\eta$. These semiclassical results are obtained from Eq.~\eqref{eq:steadystate} for $\kappa=2\gamma_h=2\pi\times 1 \,\text{MHz}$  and cooperativity parameters $C$ ranging from $4$ to $14$. Amplitude bistability shows up for $C>8$, where regions of two stable solutions (\textit{solid lines}) and one unstable solution (\textit{dashed lines}) coexist. The critical points $\eta^\pm_{\text{crit}}$ (\textit{red dots}) mark the maxima of the slope $d|\langle a_{st}\rangle|^2/d\eta$ of the input-output relation.  In the bistable regime, the cavity probability amplitude experiences a first-order phase transition at $\eta^+_{\text{crit}}$ characterized by a jump from a state of low transmission to a state of high transmission. This behavior is indicated for $C=14$ by stationary state values (\textit{blue squares}) extracted from a temporal evolution of the system under constant driving for a sufficiently long time.}
\label{fig_semicl_bistab_curves}
\end{figure}

We now analyze the validity of the semiclassical Maxwell-Bloch equations, using the CE2 and CE3 introduced in the previous section. In particular we will focus on amplitude bistability as a paradigm effect for cooperative phenomena in an open system far from equilibrium \cite{Lugiato1984}. Over the last decades amplitude bistability served as a role model for a non-equilibrium phase transition with experimental realizations in various systems \cite{Gibbs1976,Rempe1991a,Brennecke2008,Angerer2017}. To observe amplitude bistability in our model, we study the stationary transmission through a cavity-spin system under constant driving $\eta$. The transmission is proportional to the cavity probability amplitude $|\langle  a\rangle|^2$, whose stationary value $|\langle  a_{st}\rangle|^2$ can be obtained either directly by setting all time derivatives in Eqs.~\eqref{eq:MaxwellBlocha}-\eqref{eq:MaxwellBlochb} to zero or by a temporal evolution of the system for a sufficiently long time.
 
\subsection{Homogenous broadening}
\label{sec:Hom}
For simplicity, we start with the case of homogeneous coupling ($g_j=g$) and radiative decay only ($\gamma_p=0$). Moreover, we assume that the driving field is on resonance with the cavity and all individual spins ($\Delta_c=0$, $\Delta_j=0$). In the semiclassical limit, the stationary cavity probability amplitude $|\langle a_{st}\rangle|^2$ is then given by 
\begin{equation}
|\langle  a_{st}\rangle|^2\left(1+C\frac{1}{1+|\langle  a_{st}\rangle|^2/n_0}\right)^2=\eta^2/\kappa^2\,
\label{eq:steadystate}
\end{equation}
and is therefore completely characterized by the collective cooperativity parameter $C\equiv Ng^2/\kappa\gamma_h$, the photon saturation number $n_0\equiv \gamma_h^2/2g^2$, and the scaled driving amplitude $\eta/\kappa$. The stationary state equation above  gives the typical, nonlinear input-output relations presented in Fig.~\ref{fig_semicl_bistab_curves} for collective cooperativity parameters ranging from $C=4$ to $14$. Note that we increase $C$ here by increasing the individual coupling $g$ while keeping all other parameters constant.  At low driving, the enhanced cooperative emission of the ensemble leads to a stationary state of low transmission, sometimes called the lower or cooperative branch. Increasing cooperativities lead to an increasing suppression of the transmission \citep{Bonifacio1978a}. For strong driving the spins start to saturate ($\langle\sigma^-_j\rangle_{st}\to0$) and thereby decouple from the cavity (see Eq.~\eqref{eq:MaxwellBlocha}) leading to a stationary state of high transmission, which is independent of the cooperativity parameter $C$. This upper branch is therefore also called the independent-atom branch \cite{Carmichael_QO2008}. For $C>8$ the stationary state equation \eqref{eq:steadystate} exhibits regions where three solutions coexist (two stable and one unstable). This bistable region is bounded by two critical points $\eta^-_{\text{crit}}$ and $\eta^+_{\text{crit}}$, which are characterized by an infinite slope $d|\langle a_{st}\rangle|^2/d\eta$ in the transmission curve. For $C<8$ no bistability occurs and only one point of maximal but finite slope $d|\langle a_{st}\rangle|^2/d\eta$ exists. In this case there is a continuous transition from the lower to the upper branch for increasing driving. In the bistable case, by contrast, the system changes discontinuously from a state of low transmission to a state of high transmission in a first-order phase transition at the critical point $\eta^+_{\text{crit}}$.  

\begin{figure*}[t]
\includegraphics[angle=0,width=2.07\columnwidth]{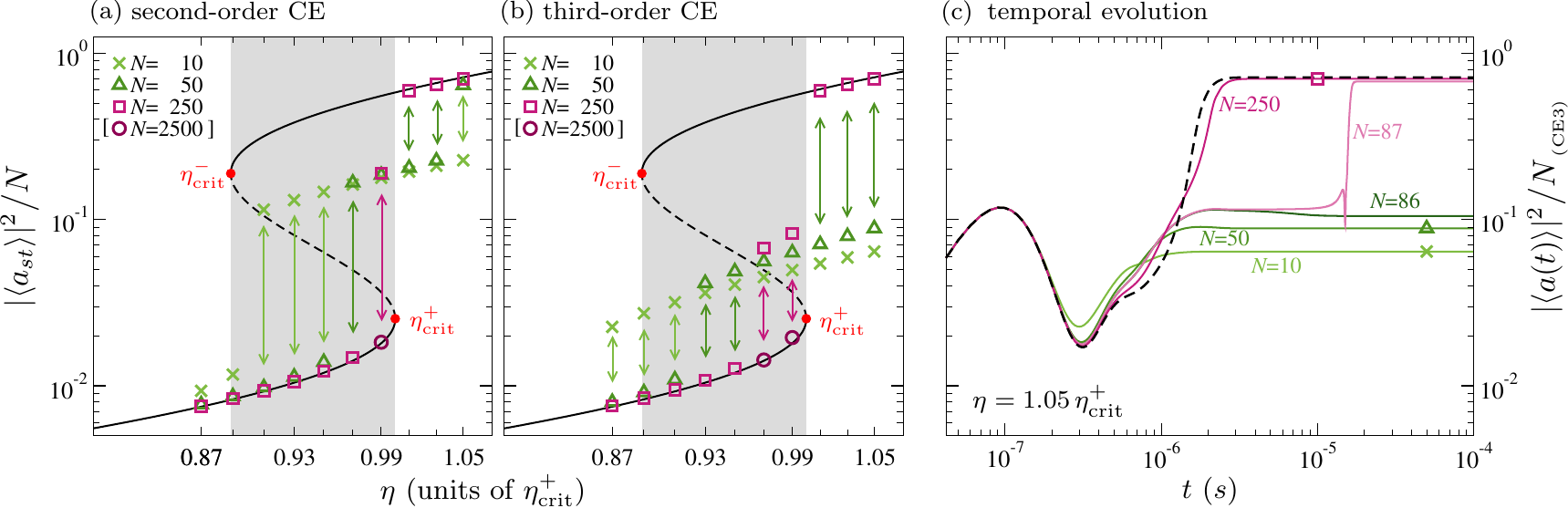}
\vspace*{-0.6cm}
\caption{Finite-size corrections to the semiclassical cavity probability amplitude for a cooperativity parameter $C=14$. (a,b) Stationary solutions for the cavity probability amplitude $|\langle a_{st}\rangle|^2$ as a function of the driving amplitude $\eta$. The semiclassical result is shown as \textit{black solid line}. Amplitude bistability is marked by two critical points $\eta^\pm_{\text{crit}}$ (\textit{red dots}) with a bistable region in between (\textit{gray area}). At well defined driving amplitudes, corrections to the semiclassical solutions are calculated using (a) the second- and (b) third-order cumulant expansion (CE) for different numbers of spins $N=10$, $50$, $250$, and $2500$.  The \textit{colored arrows} indicate the deviations from the corresponding semiclassical transmission curve. (c) Temporal evolution of the cavity probability amplitude $|\langle a(t)\rangle|^2_{\text{(CE3)}}$ using the third-order cumulant expansion (CE) and for the driving strength $\eta=1.05\,\eta^+_{\text{crit}}$. The initial conditions are chosen such that at $t=0$ the spin ensemble is unexcited and the cavity is empty. Results are shown for increasing numbers of spins, $N=10$, $50$, $86$, $87$, and $250$. The semiclassical solution is shown as \textit{black dashed line}. The \textit{colored symbols} indicate the stationary states shown in (b).}
\label{fig_bistab_curves}
\end{figure*}

In the following we are interested in the validity of the semiclassical solution in the vicinity of the critical point $\eta^+_{\text{crit}}$ for a finite number of spins. Note that, under the trivial scaling given by Eq.~\eqref{eq:scaling}, the semiclassical stationary state equation \eqref{eq:steadystate} is independent of the number of spins, $N$. The correlations $\langle\sigma_j^z a\rangle_c$ and $\langle\sigma_j^- a^\dag\rangle_c $, however, can lead to significant deviations from this trivial scaling \cite{Dombi2013}, as displayed in Fig.~\ref{fig_bistab_curves}. Here we present the impact of the quantum corrections on the stationary cavity probability amplitude $|\langle a_{st}\rangle|^2$ for a collective cooperativity parameter of $C=14$ (with the tendency described in the following being similar for all $C$). Typical numerical results using the CE2 and CE3 are demonstrated on the example of ten different driving strengths, which are chosen such that we probe the stationary states on the lower transmission branch of the bistable regime as well as on the upper transmission branch above the critical point $\eta^+_{\text{crit}}$.

Our results in Fig.~\ref{fig_bistab_curves} show that, for small ensembles, the stationary transmission calculated by means of the CE2 and CE3 deviates significantly from the semiclassical solution even outside the bistable region. As expected, increasing the number of spins restores the semiclassical results, since the quantum fluctuations decrease as $1/N$ \cite{Dombi2013}. However, the actual number of spins needed for the CE to agree well with the semiclassical solution substantially increases for driving strengths close to the critical point $\eta^+_{\text{crit}}$. Whereas the results obtained from the CE3 for spin ensembles of moderate size ($N=250$) agree reasonably well with the semiclassical solution for most driving strengths, this is not the case for the driving strengths close to the critical point of the low transmission branch $0.97\,\eta^+_{\text{crit}}$ and $0.99\,\eta^+_{\text{crit}}$, respectively. Here much larger numbers of spins ($N=2500$) are needed for the CE3 to reasonably approach the semiclassical lower transmission branch. Note that the stationary states shown in Fig.~\ref{fig_bistab_curves} are extracted from a temporal evolution of the system for sufficiently long time starting initially from an unexcited spin ensemble ($\langle\sigma_j^z\rangle=-1$, $\langle\sigma_j^-\rangle=0$) and an empty cavity ($\langle a\rangle=0$), subjected to constant driving. For this initial state the semiclassical dynamics converges towards the lower transmission branch of the bistable region and reaches the upper transmission branch only for $\eta>\eta^+_{\text{crit}}$ as indicated in Fig.~\ref{fig_semicl_bistab_curves}.

Figure~\ref{fig_bistab_curves}(c) presents the transient dynamics of the cavity probability amplitude $|\langle a(t)\rangle|^2_{_\text{\!(CE3)}}$ calculated using the CE3 for $C=14$ and driving strength $\eta=1.05\, \eta^+_{\text{crit}}$. As depicted already in Fig.~\ref{fig_bistab_curves}(b) for $N=250$ the CE3 agrees well with the semiclassical solution, whereas for small spin ensembles ($N=10$ or $50$) the CE3 tends towards a stationary state of much lower transmission than that predicted by the semiclassical equations. Interestingly, our calculations for $N=86$ and $N=87$ spins indicate, that there is an abrupt crossover from spin ensembles with large deviations from the semiclassical limit towards spin ensembles where such deviations are small. Whereas for $N\leq86$ the CE3 tends towards a stationary state of relatively low transmission, ensembles of $N\geq 87$ spins start to approach a state of high transmission, in accordance with the semiclassical solution.

\begin{figure}
\includegraphics[angle=0,width=1.\columnwidth]{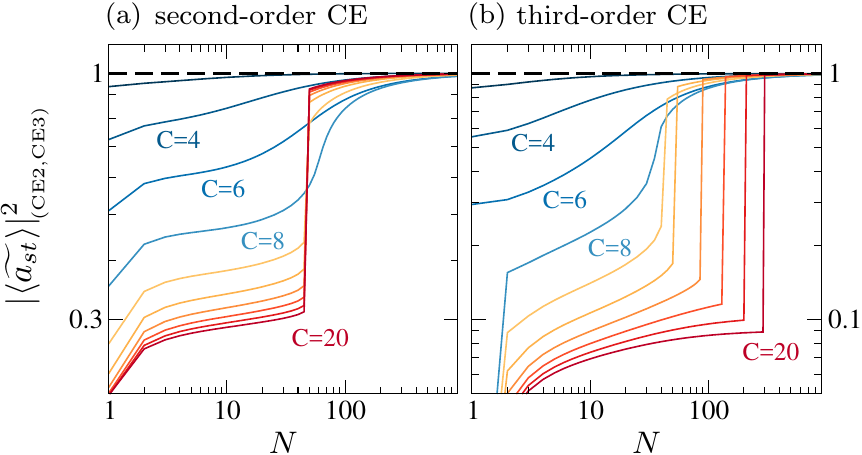}
\vspace*{-0.6cm}
\caption{Comparison of the second- and third-order cumulant expansion (CE) with the semiclassical cavity transmission for the driving strength $\eta=1.05\,\eta^+_{\text{crit}}$. The normalized stationary state solutions for the cavity probability amplitude $|\langle\widetilde{ a_{st}}\rangle|^2_{_\text{\!(CE2,3)}}=|\langle a_{st}\rangle|^2_{_\text{\!(CE2,3)}}\,\big/\,|\langle a_{st}\rangle|^2_{_\text{\!(CE1)}}$ are shown as a function of the number of spins $N$ using (a) the second- (CE2) and (b) the third-order cumulant expansions (CE3), respectively. Results are shown for cooperativity parameters $C=2$ (\textit{dark blue}) to $20$ (\textit{dark red}). $|\langle\widetilde{ a_{st}}\rangle|^2_{_\text{\!(CE2,3)}}=1$ (\textit{dashed black line}) corresponds to the semiclassical result.   }
\label{fig_normalized_transmission}
\end{figure}

%
%
To measure the validity of the semiclassical approximation for different driving strengths and cooperativity parameters, we normalize the stationary state obtained in the framework of the CE2 and CE3 by the corresponding semiclassical result (CE1),
\begin{equation}
|\langle\widetilde{ a_{st}}\rangle|^2_{_\text{\!(CE2,3)}}\equiv\frac{|\langle a_{st}\rangle|^2_{_\text{\!(CE2,3)}}}{|\langle a_{st}\rangle|^2_{_\text{\!(CE1)}}}\,,
\label{eq:normtransm}
\end{equation}
such that a value close to unity corresponds to the semiclassical regime. Figure~\ref{fig_normalized_transmission} shows the normalized cavity probability amplitude $
|\langle\widetilde{ a_{st}}\rangle|^2_{_\text{\!(CE2,3)}}$
for the driving strength $1.05\, \eta^+_{\text{crit}}$ and cooperativity parameters ranging from $C=2$ to $C=20$. Focusing at first on the CE2 solutions, we can see that the normalized cavity probability amplitude approaches the semiclassical result $|\langle\widetilde{ a_{st}}\rangle|^2_{_\text{\!(CE2)}}=1$ for increasing numbers of spins as expected from a linearized theory of fluctuations \cite{Carmichael_QO2008}. As indicated already above, our results show that for increasing cooperativity parameters the transition towards the semiclassical solution becomes more abrupt and for $C>8$ resembles a first-order phase transition at $N\approx45$. 

Coming now to the CE3 solutions, a similar tendency is observed with the only difference that the transition is shifted to larger values of $N$ for increasing cooperativities $C$. Although these results suggest that for amplitude bistability the crossover from systems of large fluctuations towards systems of small fluctuations has a discontinuous nature, care must be taken, since in the crossover region the cumulant expansion has not yet converged, i.e. the CE2 and CE3 give quite different values for $|\langle a_{st}\rangle|^2$. Including higher orders of cumulants or a full quantum mechanical treatment of the problem is therefore required to ensure the convergence to a true quantum solution in this parameter regime.

It turns out that at values of $\eta$ slightly smaller than $\eta_{\text{crit}}^+$ (at which the first-order transition obtained in the framework of the semiclassical approach occurs), other time-dependent solutions can simultaneously exist for certain numbers of spins, $N$ - a common scenario in the driven dissipative dynamics described by a set of nonlinear differential equations. Specifically, starting from the simple initial conditions mentioned above (empty cavity with unexcited spin ensemble), we end up with periodic solutions after some transient time or sometimes the overall approach even becomes numerically unstable giving rise to unphysical solutions (indicated, e.g., by unphysical values of $|\langle\sigma_k^z\rangle|>1$). To overcome this problem we vary the initial conditions for $\langle\sigma_k^z\rangle$ between $-1$ and $-0.5$ to finally find those which lie in the so-called basin of attraction for the stationary state. Originating from such initial conditions, the system eventually settles to the stationary state under study as time increases.

%
%
%

\subsection{Semiclassical-to-quantum boundary}
\label{sec:SCQboundary}
\begin{figure} 
\vspace{-0.05cm}
\includegraphics[angle=0,width=1.\columnwidth]{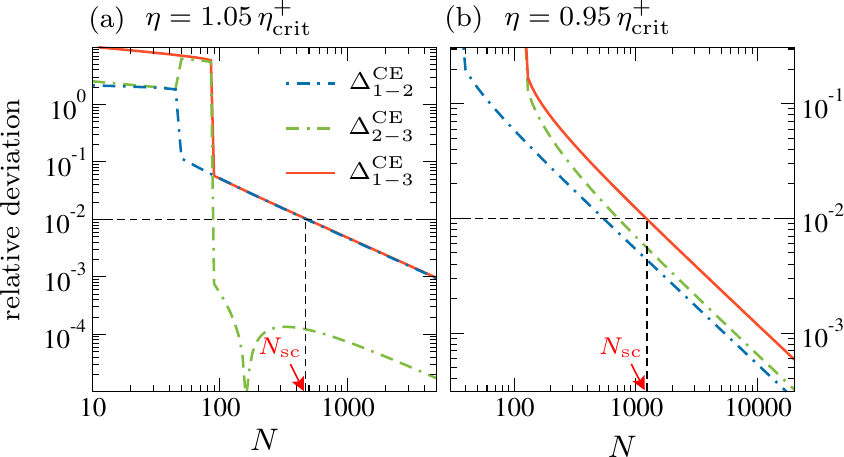}
\vspace*{-0.63cm}
\caption{Relative deviations in the cavity probability amplitude $|\langle a_{st}\rangle|^2$ calculated using the semiclassical Maxwell-Bloch equations (CE1), and  the second- (CE2), and third-order cumulant expansions (CE3). Results are shown for cooperativity $C=14$ and driving strengths (a) $\eta=1.05\,\eta^+_{\text{crit}}$ and (b) $\eta=0.95\,\eta^+_{\text{crit}}$. The relative deviations are defined as $\Delta^{\text{CE}}_{n-m}\equiv\big||\langle a_{st}\rangle|^2_{_{(n)}}\!-|\langle a_{st}\rangle|^2_{_{(m)}}\big|\big/|\langle a_{st}\rangle|^2_{_{(m)}}$, where $n$ and $m$ stand for the CE1, CE2, and CE3, respectively.  The \textit{horizontal black dashed line} indicates the threshold of convergence, which we set to be $\delta\epsilon=10^{-2}$. Only if all three relative deviations, $\Delta^{\text{CE}}_{1-2}$ (\textit{blue dashed-dotted line}), $\Delta^{\text{CE}}_{2-3}$ (\textit{green double-dashed-dotted line}), and $\Delta^{\text{CE}}_{1-3}$ (\textit{orange solid line}) are smaller than $\delta\epsilon$, the semiclassical solution for the cavity probability amplitude $|\langle a_{st}\rangle|^2$ is reliable. The minimal ensemble size that fulfils this criterion  is denoted as $N_{\text{sc}}$ (\textit{vertical black dashed line}).}
\label{fig_convergence}
\end{figure}
To avoid the difficulties in the crossover region of Fig.~\ref{fig_normalized_transmission}, we focus in the following on the results of the CE close to the semiclassical stationary states and define a criterion for the validity of the semiclassical Maxwell-Bloch equations based on the convergence of the cumulant expansion. For this purpose we define the relative deviations 
\begin{equation}
\Delta^{\text{CE}}_{n-m}\equiv\frac{\Big||\langle a_{st}\rangle|^2_{_{(n)}}\!-|\langle a_{st}\rangle|^2_{_{(m)}}\Big|}{|\langle a_{st}\rangle|^2_{_{(m)}}}\;,
\end{equation}
where $n$ and $m$ stand for the different orders of the cumulant expansion, CE1, CE2, and CE3. Figure~\ref{fig_convergence} presents the relative deviations between the first three orders of the cumulant expansion, $\Delta^{\text{CE}}_{1-2}$, $\Delta^{\text{CE}}_{2-3}$, and $\Delta^{\text{CE}}_{1-3}$, for $C=14$ and driving strengths $\eta=1.05\,\eta^+_{\text{crit}}$ and $0.95\,\eta^+_{\text{crit}}$, respectively. It turns out that the discontinuous nature of the crossover region is reflected also in the relative deviations, resulting in a rather complicated dependence on the number of spins $N$. Above the crossover regions the relative deviations decrease linearly with $1\big/N$ as expected from a linearized theory of quantum fluctuations in the small noise limit \citep{Carmichael_QO2008}. The size of the relative deviations, however, does not only strongly depend on the cooperativity $C$ and the number of spins $N$ but also on the driving strength $\eta$. Our results show that for the same $C$ and the same $N$ the relative deviations for $\eta=0.95\,\eta^+_{\text{crit}}$ are significantly larger than for $\eta=1.05\,\eta^+_{\text{crit}}$. This asymmetry will be explored in more detail below.

\begin{figure*}[t]
\includegraphics[angle=0,angle=0,width=2.07\columnwidth]{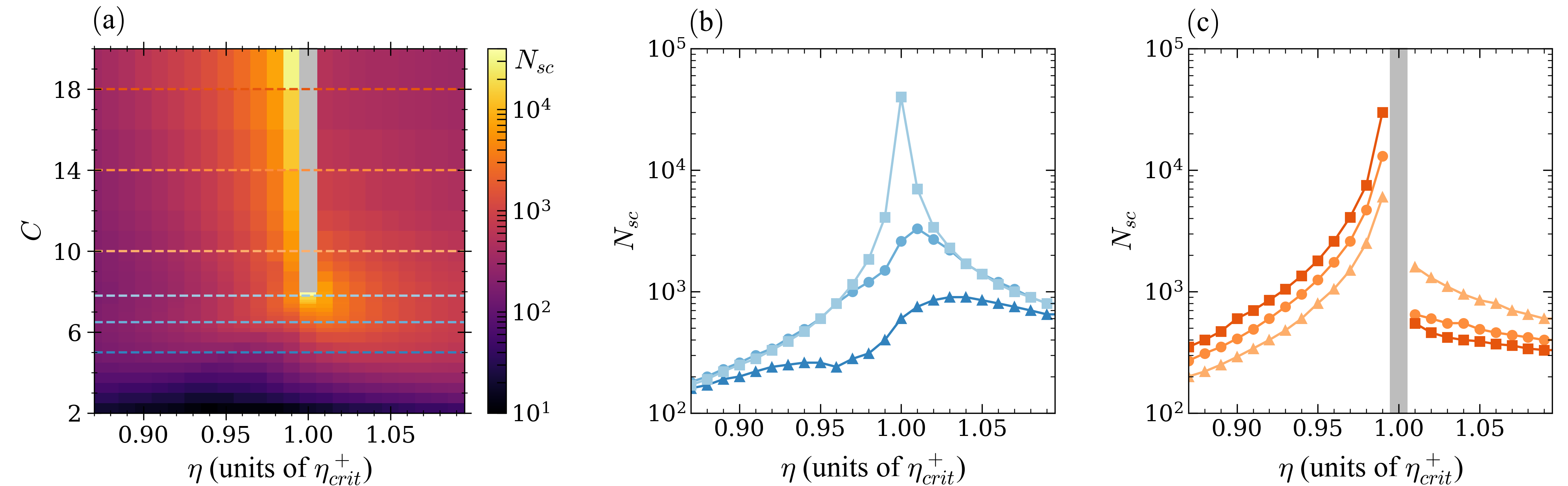}
\vspace*{-0.4cm}
\caption{Semiclassical-to-quantum boundary: Minimal number of spins $N_{\text{sc}}$ for which the cumulant expansion converges towards the semiclassical results, i.e., for which condition \eqref{eq:convergence} is fulfilled. (a) $N_{\text{sc}}$ as a function of the driving strength $\eta$ and the cooperativity parameter $C$. Note that $N_{\text{sc}}$ drastically increases in the vicinity of the critical point $\eta^+_{\text{crit}}$ for $C\geq8$ and no data are available in this region (\textit{gray bar}). The \textit{horizontal dashed lines} correspond to cooperativity parameters, which are shown separately in (b) and (c), respectively. (b) $N_{\text{sc}}$ as a function of the driving strength $\eta$ for $C=5.0$, $6.5$, and $7.8$ (\textit{dark} to \textit{light blue}). (c) $N_{\text{sc}}$ as a function of the the driving strength $\eta$ for $C=10$, $14$, and $18$ (\textit{light} to \textit{dark orange}). }
\label{fig_boundary}
\end{figure*}

In the following we define a small threshold value $\delta\epsilon=10^{-2}$, which serves as criterion for the convergence of the cumulant expansion as well as for the validity of the semiclassical solutions. We estimate the minimal number of spins, $N_{\text{sc}}$, for which all relative deviations drop below the threshold value, i.e.
\begin{equation}
\Delta^{\text{CE}}_{1-2},\;\Delta^{\text{CE}}_{2-3},\;\Delta^{\text{CE}}_{1-3}\;<\delta \epsilon\;,
\label{eq:convergence}
\end{equation}and call it a semiclassical-to-quantum boundary. Hence, for spin ensembles with $N>N_{\text{sc}}$ the semiclassical Maxwell-Bloch equations provide trustworthy results for the cavity probability amplitude $|\langle a_{st}\rangle|^2$ and all higher correlations like $\langle\sigma_j^z a\rangle_c$, $\langle\sigma_j^- a^\dag\rangle_c $, etc., give only negligible relative contributions (lower than 1\%). The vanishing influence of these quantum correlations, which is characterized by Eq.~\eqref{eq:convergence}, captures a subtle but important point of open quantum systems. While the true quantum solution of the steady state master equation $\dot{\rho}=0$ is unique, the bistability of the semiclassical regime translates into a bimodality of the corresponding quasi-probability function. The quantum system switches between these two quasistationary components giving a unique time averaged expectation value which can substantially deviate from the semiclassical solution within the bistability region \cite{Dombi2013,Casteels2017,Fink2017}. The switching time, however, strongly depends on the system size being divergent for large $N$ \cite{Casteels2017,Fink2017}. Hence, the convergence criterion provided by Eq.~\eqref{eq:convergence} implies that for $N>N_{\text{sc}}$, the cavity probability amplitude $|\langle a_{st}\rangle|^2$ calculated from a full quantum mechanical evolution of the initial state can not be distinguished from the solutions of the semiclassical Maxwell-Bloch equations on experimentally feasible timescales.

%
%

Figure~\ref{fig_boundary} shows this semiclassical-to-quantum boundary value as a function of both the cooperativity parameters $C$ and the driving strengths $\eta$. This main result of our paper demonstrates how the value of $N_{\text{sc}}$ increases close to the critical point $\eta^+_{\text{crit}}$. Exactly at $\eta=\eta^+_{\text{crit}}$ the time the systems needs to reach its stationary state for $C\geq8$ diverges due to the effect of critical slowing down \cite{Bonifacio1979} and data points are therefore omitted for these parameters. For parameter regions where no bistability occurs, i.e. for $C< 8$, there is no sharp distinction between the lower and upper transmission branch. Here the value of the semiclassical-to-quantum boundary $N_{\text{sc}}$ has its maximum at driving strengths slightly above the critical driving strength $\eta^+_{\text{crit}}$ and starts to peak at $\eta=\eta^+_{\text{crit}}$ only as $C$ approaches the threshold value of bistability. As can be seen in Fig.~\ref{fig_boundary}(b) for the cooperativity parameter $C=5$ the value of $N_{\text{sc}}$ above $\eta^+_{\text{crit}}$ is significantly larger than below that driving strength. This asymmetry with respect to $\eta^+_{\text{crit}}$ becomes less pronounced but is still present for increasing cooperativities up to $C\leq8$. The peak in $N_{\text{sc}}$ for $C=7.8$ at the critical driving is the precursor of the emergence of a first-order phase transition and the effect of bistability, which emerges for cooperativity values above $C=8$.

It is worth noting that for $C>8$ the semiclassical-to-quantum boundary as defined in Eq.~\eqref{eq:convergence} behaves qualitatively differently for the lower transmission branch as compared to the upper transmission branch. As we approach the critical point $\eta^+_{\text{crit}}$ from below by analyzing stationary states disposed on the lower transmission branch, the value $N_{\text{sc}}$ progressively increases and eventually diverges exactly at $\eta^+_{\text{crit}}$, where the saddle-node bifurcation occurs [curves from the left with respect to the gray bar of Fig.~\ref{fig_boundary}(c)]. In contrast, the only stable solutions above $\eta^+_{\text{crit}}$ are those which are located on the upper transmission branch which lie, however, far away in phase space from the saddle-node bifurcation at $\eta^+_{\text{crit}}$. Therefore when approaching $\eta^+_{\text{crit}}$ from above, the value of $N_{\text{sc}}$ exhibits no divergence and is significantly smaller than below $\eta^+_{\text{crit}}$. 

Another interesting tendency seen in Fig.~\ref{fig_boundary}(c) is that for $\eta<\eta^+_{\text{crit}}$ (lower transmission branch) $N_{\text{sc}}$ increases for increasing cooperativity parameters $C$. In contrast, for $\eta>\eta^+_{\text{crit}}$ (upper transmission branch) increasing cooperativity parameters $C$ lead to a decrease of $N_{\text{sc}}$. Our findings suggest that close to the critical point at the lower transmission branch even very large ensembles of up to $\sim10^4$ spins can show behavior that goes beyond the semiclassical description. Note that for large cooperativities the differences in $N_{\text{sc}}$ for driving strengths within and outside the bistable region become very large. Whereas for $C=18$ and $\eta=1.01\,\eta^+_{\text{crit}}$ the semiclassical result agrees well (1\% deviation) with the CE2 and CE3  already for ensembles of $N_{\text{sc}}\approx500$ spins, the corresponding value grows to $N_{\text{sc}}\approx3\times10^4$ for $\eta=0.99\,\eta^+_{\text{crit}}$.
This can be explained by very large quantum fluctuations near the critical point, which destabilize one of the two semiclassical basins of attraction. Note that the asymmetry between the lower and upper transmission branches encountered in Fig.~\ref{fig_boundary}(c) results from our choice to probe the critical point $\eta^+_\text{crit}$. We anticipate a reversed role of the two transmission branches when probing the stationary states of the upper transmission branch close to the critical point $\eta^-_\text{crit}$. 

\subsection{Inhomogeneous broadening}
\label{sec:inHom}
\begin{figure*}[]
\includegraphics[angle=0,angle=0,width=2.07\columnwidth]{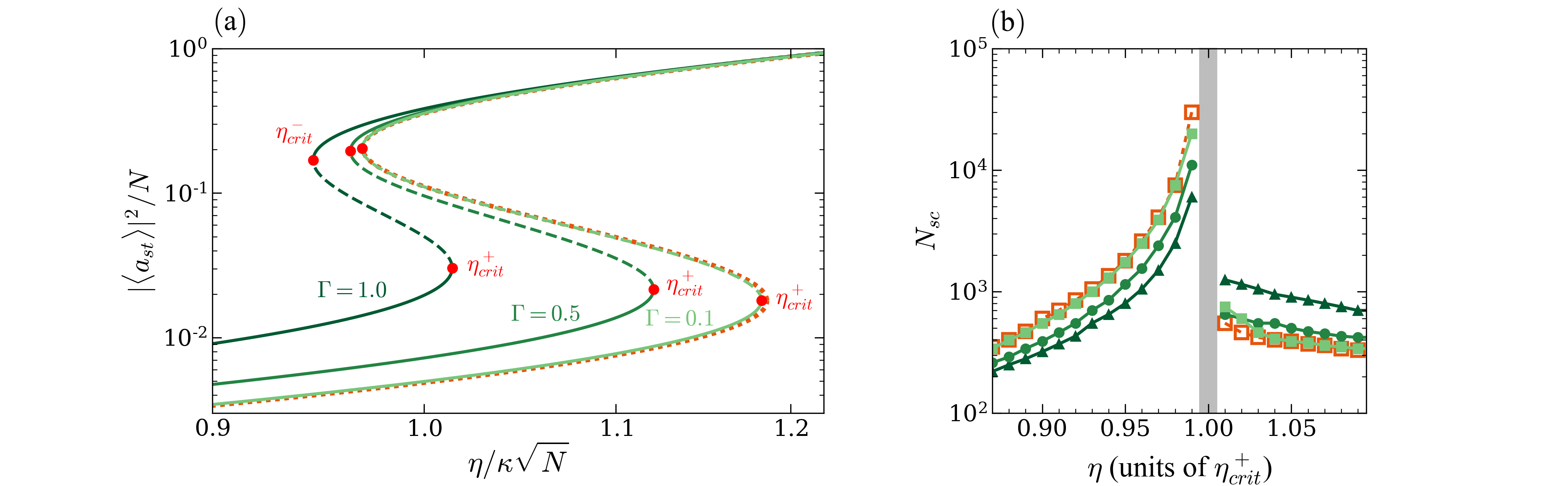}
\vspace*{-0.4cm}
\caption{Semiclassical-to-quantum boundary including inhomogeneous broadening: (a) Semiclassical stationary states for the cavity probability amplitude $|\langle a_{st}\rangle|^2$ as a function of the driving amplitude $\eta$. Parameters are the same as in Fig.~\ref{fig_bistab_curves}. Results are shown for Gaussian spin distributions with a full width at half maximum of $\Gamma=0.1$, $0.5$, and $1.0$ MHz, corresponding to a collective cooperativity of $C\approx 17.9$, $15.8$, and $12.7$, respectively (\textit{light} to \textit{dark green}). Results without inhomogeneous broadening, corresponding to $C=18$, are presented for comparison purposes (\textit{dotted orange line}).  (b) Semiclassical-to-quantum boundary $N_{\text{sc}}$ as a function of the the driving strength $\eta$ for $\Gamma=0.1$, $0.5$, and $1.0$ MHz (\textit{light} to \textit{dark green}), and no inhomogeneous broadening (\textit{orange}). }
\label{fig_inhom}
\end{figure*}

In the following we extend our investigations to inhomogeneously broadened spin ensembles and study how the broadening of the spin ensemble alters the previously defined minimal number of spins $N_{\text{sc}}$ that is required for the validity of the semiclassical Maxwell-Bloch equations. We therefore relax the condition $\Delta_j=0$ and allow for a Gaussian frequency distribution of the individual spins. This scenario becomes computationally much more demanding, since now the hierarchy of equations (summarized in the Appendix) has to be solved for each spin frequency $\Delta_j$ individually. Numerically we split the spin ensemble into $L$ equidistantly spaced frequency clusters $\Delta_\mu$, where the index $\mu$ runs from $1$ to $L$ and each frequency cluster is filled up with $M_\mu$ spins. After this procedure, the equations (\ref{eq: 1})-(\ref{eq: 25}), representing the CE3, in total are $13L^2+L(L+1)/2+23L+9$ first-order ordinary differential equations, which can be solved for moderate values of $L$. For our calculations we chose $L=51$ and distributed the spins following a Gaussian distribution 
\begin{equation}
M_\mu=\frac{N}{K}\,e^{-4\ln(2)\frac{\Delta_\mu^2}{\Gamma^2}}\,,
\end{equation} 
with $K=\sum_{\nu=1}^Le^{-4\ln(2)\frac{\Delta_\nu^2}{\Gamma^2}}$ being a normalization constant such that $\sum_{\mu=1}^{L}M_\mu=N$.

In Fig.~\ref{fig_inhom} we present results for Gaussian spin distributions with three different full widths at half maximum, $\Gamma=0.1$, $0.5$, and $1.0$ MHz. Note that an increase of the width $\Gamma$ leads to a decrease of the collective cooperativity
\begin{equation}
C=\frac{g^2}{\kappa\gamma_h}\sum_{\mu=1}^LM_\mu\frac{1}{1+\Delta_\mu^2/\gamma_h^2},
\end{equation}
with all other parameters kept constant. This drop of the collective cooperativity for increasing widths of the distribution can be observed in Fig.~\ref{fig_inhom}(a), where we show the semiclassical bistability curves of the inhomogeneously broadened spin ensembles in comparison with the unbroadened case (corresponding to a collective cooperativity of $C=18$). The spin distributions with a full width at half maximum of $\Gamma=0.1$, $0.5$, and $1.0$ MHz, then corresponds to a collective cooperativity parameters of $C\approx17.9$, $15.8$, and $12.7$, respectively.

The minimal number of spins, $N_{\text{sc}}$, for which the cumulant expansion converges towards the semiclassical results is presented in Fig.~\ref{fig_inhom}(b). A comparison with Fig.~\ref{fig_boundary}(c) indicates that the change in the semiclassical-to-quantum boundary due to the inhomogeneous broadening can be well understood in the way the collective cooperativity parameter $C$ changes with broadening. Our results for the inhomogeneously broadened spin distributions therefore confirm our earlier findings that even very large spin ensembles of about $10^4$ spins can show non-semiclassical behavior close to the critical point of bistability.

\section{Conclusions}
\label{sec:conclusions}
We have studied in detail the route towards the semiclassical limit for a dissipative spin-cavity system driven close to the critical point of amplitude bistability. In particular we analyzed the validity of the semiclassical Maxwell-Bloch equations close to the critical point following the transition from the lower to the upper transmission branch of amplitude bistability for varying cooperativities $C$ and for different numbers of spins $N$. We numerically solved the nonlinear sets of equations resulting from a second- (CE2) and third-order cumulant expansion (CE3) and compared the results with the semiclassical stationary solution for the cavity probability amplitude. Based on the convergence of the cumulant expansion towards the semiclassical results, we defined a criterion for the reliability of the Maxwell-Bloch equations and determined the minimal number of spins $N_\text{sc}$ necessary to ensure the validity of the semiclassical approximation. 

Our results reveal that not only the distance to but also the way of approaching the critical point is a crucial factor which strongly influences the validity of the semiclassical equations. More specifically, we disclose that the large quantum fluctuations inside the bistable region lead to very large values of $N_\text{sc}$ in the proximity of the critical point of the lower transmission branch. Remarkably, here even very large spin ensembles of up to $\approx10^4$ spins can feature deviations from the semiclassical cavity probability amplitude. Our results therefore suggest that a spin ensemble of the same size can behave semiclassically or quantum mechanically depending not only on the system parameters but also on the proximity of critical points and the way of approaching them.
\section*{Acknowledgements}
We would like to thank H. Dhar, U. Hohenester, and P. Rabl for helpful discussions and the European Commission under Project No.~NHQWAVE (MSCA-RISE 691209) for support. S.R. and M.Z. acknowledge support by the Austrian Science Fund (FWF) through Project No. F49-P10 (SFB NextLite) and the Doctoral Program CoQuS (W1210). Some of the computational results presented have been achieved using the Vienna Scientific Cluster (VSC).
\section*{Appendix: Hierarchy of coupled equations}
\renewcommand{\theequation}{\Alph{section}.\arabic{equation}}
\setcounter{section}{1}
\setcounter{equation}{0}
In order to employ the CE2 and CE3 as described in Sec.~\ref{sec:CE} the equations of motions have to be generated also for second- and third-order expectation values. The derivation of these equations is straightforward  but the arising expressions soon become unwieldy for higher orders of expectation values. Since the resulting equations, which enable an accurate description of spin-cavity systems including the effect of inhomogeneous broadening, constitute a critical part of our paper, we provide them explicitly below:

{\it First-order expectation values:}
\begin{flalign} 
\frac{d}{dt}\braket{a}&=-(\kappa+i\,\Delta_c)\braket{a}-i\,\sum_{k=1}^Ng_k\braket{\sigma_k^-}+\eta&
\label{eq: 1}
\end{flalign} 
\begin{flalign} 
\frac{d}{dt}\braket{\sigma_k^-}&=-(\gamma_h+2\gamma_p+i\,\Delta_k)\braket{\sigma_k^-}+i\,g_k\braket{\sigma_k^za}&
\label{eq: 2}
\end{flalign}  
\begin{flalign} 
\frac{d}{dt}\braket{\sigma_k^z}&=-2\gamma_h(\braket{\sigma_k^z}+1)+2i\,g_k(\,\braket{\sigma_k^-a^\dag}-\braket{\sigma_k^-a^\dag}^*\,)&
\label{eq: 3}
\end{flalign}  
%
{\it Second-order expectation values:}
\begin{flalign} 
\frac{d}{dt}\braket{\sigma_k^za}=&-(\kappa+2\gamma_h+i\,\Delta_c)\braket{\sigma_k^za}-2\gamma_h\braket{a}\notag\\
&+\eta\braket{\sigma_k^z}-i\,\sum_{\substack{j=1\\j\neq k}}^Ng_j\braket{\sigma_k^z\sigma_j^-}+ig_k\braket{\sigma_k^-}\notag\\
&+2i\,g_k(\,\braket{\sigma_k^-a^\dag a}-\braket{\sigma_k^-a^\dag a^\dag}^*\,)&
\label{eq: 4}
\end{flalign} 
%
\begin{flalign} 
\frac{d}{dt}\braket{\sigma_k^z\sigma_j^-}\underset{j\neq k}{=}&-(3\gamma_h+2\gamma_p+i\,\Delta_j)\braket{\sigma_k^z\sigma_j^-}-2\gamma_h\braket{\sigma_j^-}
\notag\\
&+i\,g_j\braket{\sigma_k^z\sigma_j^z a}+ 2i\,g_k(\,\braket{\sigma_k^-\sigma_j^-a^\dag}\notag\\
&-\braket{\sigma_k^+\sigma_j^- a}\,)&
\label{eq: 5}
\end{flalign} 
\begin{flalign} 
\frac{d}{dt}\braket{\sigma_k^-a^\dagger}=&-(\kappa+\gamma_h+2\gamma_p+i\,(\Delta_k-\Delta_c)) \braket{\sigma_k^-a^\dagger}\notag\\
&+\eta\,\braket{\sigma_k^-}+i\,\sum_{\substack{j=1\\j\neq k}}^Ng_j\braket{\sigma_j^+\sigma_k^-}\notag\\[-3mm]
&+i\,\frac{g_k}{2}\,(\braket{\sigma_k^z}+1)+i\,g_k\braket{\sigma_k^za^\dag a}&
\label{eq: 6}
\end{flalign} 
\begin{flalign} 
\frac{d}{dt}\braket{\sigma_k^+\sigma_j^-}\underset{j\neq k}{=}&-(2\gamma_h+4\gamma_p+i\,(\Delta_j-\Delta_k)) \braket{\sigma_k^+\sigma_j^-}\notag\\
&-i\,g_k\braket{\sigma_k^z\sigma_j^-a^\dag}
+i\,g_j\braket{\sigma_j^z\sigma_k^-a^\dag}^*&
\label{eq: 7}
\end{flalign} 
\begin{flalign} 
\frac{d}{dt}\braket{\sigma_k^-a}=&-(\kappa+\gamma_h+2\gamma_p+i\,(\Delta_k+\Delta_c)) \braket{\sigma_k^-a}\notag\\
&+\eta\braket{\sigma_k^-}-i\,\sum_{\substack{j=1\\j\neq k}}^Ng_j\braket{\sigma_k^-\sigma_j^-}+i\,g_k\braket{\sigma_k^zaa}&
\label{eq: 8}
\end{flalign} 
\begin{flalign} 
\frac{d}{dt}\braket{a^\dagger a^\dagger}=&-2(\kappa-i\,\Delta_c)\braket{a^\dagger a^\dagger} +2i\,\sum_{k=1}^Ng_k\braket{\sigma_k^-a}^*\notag\\
&+2\eta\braket{a}^*&
\label{eq: 9}
\end{flalign} 
\begin{flalign} 
\frac{d}{dt}\braket{a^\dagger a}=&-2\kappa\braket{a^\dagger a}-i\sum_{k=1}^Ng_k(\braket{\sigma_k^-a^\dagger}-\braket{\sigma_k^-a^\dagger}^*)\notag\\
&+\eta\;( \braket{a}+\braket{a}^*)&
\label{eq: 10}
\end{flalign} 
\begin{flalign} 
\frac{d}{dt}\braket{\sigma_k^z\sigma_j^z}\underset{j\neq k}{=}&-2\gamma_h(\braket{\sigma_k^z}+\braket{\sigma_k^z\sigma_j^z}
+\braket{\sigma_j^z}+\braket{\sigma_j^z\sigma_k^z}) \notag\\
&+2i\,g_k(\,\braket{\sigma_j^z\sigma_k^-a^\dag}-\braket{\sigma_j^z\sigma_k^-a^\dag}^*\,)
\notag\\
&+2i\,g_j(\,\braket{\sigma_k^z\sigma_j^-a^\dag}-\braket{\sigma_k^z\sigma_j^-a^\dag}^*\,)&
\label{eq: 11}
\end{flalign} 
\begin{flalign} 
\frac{d}{dt}\braket{\sigma_k^-\sigma_j^-}\underset{j\neq k}{=}&-(2\gamma_h+4\gamma_p+i\,(\Delta_j+\Delta_k))\braket{\sigma_k^-\sigma_j^-} \notag\\
&+i\,g_k\braket{\sigma_k^z\sigma_j^-a}+i\,g_j\braket{\sigma_j^z\sigma_k^-a}&
\label{eq: 12}
\end{flalign} 
%
%
{\it Third-order expectation values:}
\begin{flalign} 
\frac{d}{dt}\braket{\sigma_k^za^\dag a}=&-2(\kappa+\gamma_h)\braket{\sigma_k^za^\dag a}-2\gamma_h\braket{a^\dag a}+\eta(\braket{\sigma_k^za}\notag\\
&+\braket{\sigma_k^za}\!\!^*)-i\sum_{\substack{j=1\\j\neq k}}^Ng_j(\braket{\sigma_k^z\sigma_j^-a^\dag}\!-\!\braket{\sigma_k^z\sigma_j^-a^\dag}\!\!^*)\notag\\
&+i\,g_k(\braket{\sigma_k^-a^\dag}-\braket{\sigma_k^-a^\dag}\!^*)\notag\\
&+2i\,g_k(\braket{\sigma_k^-a^\dag a^\dag a}-\braket{\sigma_k^-a^\dag a^\dag a}\!\!^*)&
\label{eq: 13}
\end{flalign} 
\begin{flalign} 
\frac{d}{dt}\braket{\sigma_k^-a^\dag a}=&-(2(\kappa+\gamma_p)+\gamma_h+i\Delta_k)\braket{\sigma_k^-a^\dag a}\notag\\
&+\eta(\braket{\sigma_k^-a^\dag}+\braket{\sigma_k^-a})+i\,g_k\braket{\sigma_k^za^\dag a a}&\notag\\
&+i\,\sum_{\substack{j=1\\j\neq k}}^Ng_j(\braket{\sigma_j^+\sigma_k^- a}-\braket{\sigma_k^-\sigma_j^- a^\dag})\notag\\
&+i\,\frac{g_k}{2}(\braket{\sigma_k^z a}+\braket{a})&
\label{eq: 14}
\end{flalign} 
\begin{flalign} 
\frac{d}{dt}\braket{\sigma_k^-a^\dag a^\dag}=&-(2(\kappa+\gamma_p)+\gamma_h\!+\!i(\Delta_k\!-\!2\Delta_c))\braket{\sigma_k^-a^\dag a^\dag}\notag\\
&+2\eta\braket{\sigma_k^-a^\dag}+2i\,\sum_{\substack{j=1\\j\neq k}}^Ng_j\braket{\sigma_j^+\sigma_k^- a^\dag}\notag\\
&+i\,g_k(\braket{\sigma_k^z a}^*+\braket{a}^*)
+i\,g_k\braket{\sigma_k^za^\dag a^\dag a}&
\label{eq: 15}
\end{flalign} 
\begin{flalign} 
\frac{d}{dt}\braket{\sigma_k^za a}=&-2(\kappa+\gamma_h+i\Delta_c)\braket{\sigma_k^za a}-2\gamma_h\braket{a^\dag a^\dag}^*\notag\\
&+2\eta\braket{\sigma_k^za}+2i\,g_k\braket{\sigma_k^- a}-2i\sum_{\substack{j=1\\j\neq k}}^Ng_j\!\braket{\sigma_k^z\sigma_j^- a}\notag\\
&
+2i\,g_k(\braket{\sigma_k^-a^\dag a a}-\braket{\sigma_k^+a a a})&
\label{eq: 16}
\end{flalign} 
\begin{flalign} 
\frac{d}{dt}\braket{\sigma_k^-a a}=&-(2(\kappa+\gamma_p)+\gamma_h+i(\Delta_k+2\Delta_c))\braket{\sigma_k^-a a}\notag\\
&+2\eta\braket{\sigma_k^-a}-2i\,\sum_{\substack{j=1\\j\neq k}}^Ng_j\braket{\sigma_k^-\sigma_j^- a}\notag\\
&+i\,g_k\braket{\sigma_k^za a a}&
\label{eq: 17}
\end{flalign} 
\begin{flalign} 
\frac{d}{dt}\braket{a^\dagger a a}=&-(3\kappa+i\,\Delta_c)\braket{a^\dagger a a} -2i\,\sum_{k=1}^Ng_k\braket{\sigma_k^-a^\dag a}\notag\\
&+i\,\sum_{k=1}^Ng_k\braket{\sigma_k^-a^\dag a^\dag}^*+2\eta\braket{a^\dag a}+\eta\braket{a^\dag a^\dag}\!\!^*&
\label{eq: 18}
\end{flalign} 
\begin{flalign} 
\frac{d}{dt}\braket{a a a}=&-3(\kappa+i\,\Delta_c)\braket{a a a} -3i\,\sum_{k=1}^Ng_k\braket{\sigma_k^-a a}\notag\\
&+3\eta\braket{a^\dag a^\dag}\!\!^*&
\label{eq: 19}
\end{flalign} 
\begin{flalign} 
\frac{d}{dt}\braket{\sigma_k^z\sigma_j^z a}\underset{j\neq k}{=}&\!-(\kappa\!+\!i\Delta_c)\braket{\sigma_k^z\sigma_j^z a}\!-2\gamma_h(\braket{\sigma_k^za}\!+\!\braket{\sigma_k^z\sigma_j^za}\notag\\
&
+\braket{\sigma_j^za}+\braket{\sigma_j^z\sigma_k^za})+2i(g_k\braket{\sigma_j^z\sigma_k^-a^\dag a}\notag\\
&+g_j\braket{\sigma_k^z\sigma_j^-a^\dag a}-g_k\braket{\sigma_j^z\sigma_k^+a a}\notag\\
&-g_j\braket{\sigma_k^z\sigma_j^+a a})-i\sum_{\substack{m\,=\,1_{}\\m\neq k,j}}^Ng_m\braket{\sigma_k^z\sigma_j^z \sigma_m^-}\notag\\
&+i\,g_k\braket{\sigma_j^z\sigma_k^-}+i\,g_j\braket{\sigma_k^z\sigma_j^-}&
\label{eq: 20}
\end{flalign} 
\begin{flalign} 
\frac{d}{dt}\braket{\sigma_k^-\sigma_j^- a^\dag}\underset{j\neq k}{=}&-(\kappa+2\gamma_h+4\gamma_p\notag\\
&+i(\Delta_k+\Delta_j-\Delta_c))\braket{\sigma_k^-\sigma_j^- a^\dag}\notag\\
&+\eta\braket{\sigma_k^-\sigma_j^-}+i\sum_{\substack{m\,=\,1_{}\\m\neq k,j}}^Ng_m\braket{\sigma_m^+\sigma_k^- \sigma_j^-}\notag\\
&+i\frac{g_k}{2}(\braket{\sigma_j^-}+\braket{\sigma_k^z\sigma_j^-})+i\frac{g_j}{2}(\braket{\sigma_k^-}\notag\\
&+\braket{\sigma_j^z\sigma_k^-})+i\,g_k\braket{\sigma_k^z\sigma_j^-a^\dag a}\notag\\
&+i\,g_j\braket{\sigma_j^z\sigma_k^-a^\dag a}&
\label{eq: 21}
\end{flalign} 
\begin{flalign} 
\frac{d}{dt}\braket{\sigma_k^+\sigma_j^- a}\underset{j\neq k}{=}&-(\kappa+2\gamma_h+4\gamma_p\notag\\
&+i(\Delta_j\!-\!\Delta_k\!+\!\Delta_c))\braket{\sigma_k^+\sigma_j^- a}+\eta\braket{\sigma_k^+\sigma_j^-}&\notag\\
&-i\sum_{\substack{m\,=\,1_{}\\m\neq k,j}}^Ng_m\braket{\sigma_k^+\sigma_j^- \sigma_m^-}-i\frac{g_k}{2}(\braket{\sigma_j^-}\notag\\
&+\braket{\sigma_k^z\sigma_j^-})-i\,g_k\braket{\sigma_k^z\sigma_j^-a^\dag a}\notag\\
&+i\,g_j\braket{\sigma_k^+\sigma_j^za a}&
\label{eq: 22}
\end{flalign} 
\begin{flalign} 
\frac{d}{dt}\braket{\sigma_k^z\sigma_j^- a^\dag}\underset{j\neq k}{=}&-(\kappa+3\gamma_h+2\gamma_p+i(\Delta_j\!-\!\Delta_c))\braket{\sigma_k^z\sigma_j^- a^\dag}\notag\\
&-2\gamma_h\braket{\sigma_j^-a^\dag}+\eta\braket{\sigma_k^z\sigma_j^-}&\notag\\
&+i\sum_{\substack{m\,=\,1_{}\\m\neq k,j}}^Ng_m\braket{\sigma_m^+\sigma_k^z\sigma_j^-}-i\,g_k\braket{\sigma_k^+\sigma_j^-}\notag\\
&+i\,\frac{g_j}{2}(\braket{\sigma_k^z}+\braket{\sigma_k^z\sigma_j^z})+i\,g_j\braket{\sigma_k^z\sigma_j^za^\dag a}\notag\\
&+2i\,g_k(\braket{\sigma_k^-\sigma_j^-a^\dag a^\dag}-\braket{\sigma_k^+\sigma_j^-a^\dag a})&
\label{eq: 23}
\end{flalign} 
\begin{flalign} 
\frac{d}{dt}\braket{\sigma_k^z\sigma_j^- a}\underset{j\neq k}{=}&-(\kappa+3\gamma_h+2\gamma_p+i(\Delta_j+\Delta_c))\braket{\sigma_k^z\sigma_j^- a}\notag\\
&-2\gamma_h\braket{\sigma_j^-a}+\eta\braket{\sigma_k^z\sigma_j^-}&\notag\\
&-i\sum_{\substack{m\,=\,1_{}\\m\neq k,j}}^Ng_m\braket{\sigma_k^z\sigma_j^-\sigma_m^-}+i\,g_k\braket{\sigma_k^-\sigma_j^-}\notag\\
&+i\,g_j\braket{\sigma_k^z\sigma_j^z a a}+2i\,g_k(\braket{\sigma_k^-\sigma_j^-a^\dag a}\notag\\
&-\braket{\sigma_k^+\sigma_j^-a a})&
\label{eq: 24}
\end{flalign} 
\begin{flalign} 
\frac{d}{dt}\braket{\sigma_k^-\sigma_j^- a}\underset{j\neq k}{=}&-(\kappa+2\gamma_h+4\gamma_p\notag\\
&+i(\Delta_k+\Delta_j+\Delta_c))\braket{\sigma_k^-\sigma_j^- a}\notag\\
&+\eta\braket{\sigma_k^-\sigma_j^-}-i\sum_{\substack{m\,=\,1_{}\\m\neq k,j}}^Ng_m\braket{\sigma_k^-\sigma_j^-\sigma_m^-}\notag\\
&+i\,g_k\braket{\sigma_k^z\sigma_j^- a a}+i\,g_j\braket{\sigma_j^z\sigma_k^- a a}&
\label{eq: 25}
\end{flalign}

\end{document}